\documentclass[12pt]{article}
\usepackage{amsmath,amsfonts}
\advance \textheight by \topskip
\newcommand{\beq}{\begin{equation}}
\newcommand{\bee}{\end{equation}}
\newcommand{\beqa}{\begin{eqnarray}}
\newcommand{\eeqa}{\end{eqnarray}}
\newcommand{\noi}{\noindent}
\newcommand{\e}{\varepsilon}
\newcommand{\ttA}{{\tilde {\tilde A}}}
\newcommand{\tA}{\tilde  A}

\newcommand{\tB}{\tilde  B}

\newcommand{\hA}{\hat  A}
\newcommand{\hhA}{{\hat {\hat A}}}
\newcommand{\hB}{\hat  B}
\newcommand{\hhB}{{\hat {\hat B}}}
\def\ZZ{\hbox{\it Z\hskip -4.pt Z}}

\begin{document}
\title{Non-trivial extension of the Poincar\'e algebra
for antisymmetric gauge fields
}
\author{    
{\sf   G. Moultaka} \thanks{e-mail:
moultaka@lpm.univ-montp2.fr}$\,\,$${}^{a},$
{\sf  M. Rausch de Traubenberg }\thanks{e-mail:
rausch@lpt1.u-strasbg.fr}$\,\,$${}^{b}$
 and
{\sf A. Tanasa}\thanks{e-mail:
atanasa@lpt1.u-strasbg.fr}$\,\,$$^{c,b}$\\
\\
{\small ${}^{a}${\it Laboratoire de Physique 
Math\'ematique et Th\'eorique, CNRS UMR 5825, 
Universit\'e Montpellier II,}}\\
{\small {\it Place E. Bataillon, 34095 Montpellier,
France}}\\
{\small ${}^{b}${\it
Laboratoire de Physique Th\'eorique, CNRS UMR  7085,
Universit\'e Louis Pasteur}}\\
{\small {\it  3 rue de 
l'Universit\'e, 67084 Strasbourg, France}}\\ 
{\small ${}^{c}${\it
Laboratoire de Math\'ematique et Applications,
Universit\'e de Haute Alsace }}\\
{\small {\it  Facult\'e des Sciences et Techniques, 4 rue des 
fr\`eres Lumi\`eres 68093  Mulhouse, France}} 
}
\date{}
\maketitle

%
%

\maketitle
\begin{abstract}
We investigate a non-trivial  extension of the $D-$dimensional 
Poincar\'e algebra. Matrix representations are obtained.
The bosonic multiplets contain antisymmetric
tensor fields. It turns out that this symmetry acts in
a natural geometric way on these $p-$forms. Some field
theoretical aspects of this symmetry are studied and invariant
Lagrangians are explicitly given.
\end{abstract}


keywords: algebraic methods, extension of the Poincar\'e algebra, $p-$forms,
field theory


\section{Introduction}    
The usual electromagnetic gauge field admits a generalisation
in terms of antisymmetric $p$th order tensors or $p-$forms. The
revival of interest for  the $p-$forms is  mainly due to there appearance in
supergravity or string theory. Furthermore, the $p$th antisymmetric
gauge fields naturally couple to $(p-1)-$dimensional extended objects.
However, the $p-$forms  are known to be relatively rigid in the
sense that there is a few number of consistent interactions for
them \cite{h}. Despite these restrictions  $p-$forms
may present interesting symmetry properties, as it is the
case, for instance, with the  duality transformations between  types
IIA and  IIB string theories \cite{A-B}. In this paper we investigate
a new possible symmetry among $p-$forms. This symmetry
turns out to be a non-trivial extension
of the Poincar\'e algebra in $D-$dimensional  space-time, 
an extension different from  supersymmetry.

By  definition this symmetry 
is not in contradiction with the no-go theorems \cite{c-h}
because its underlying algebraic structure is neither
a Lie algebra nor a Lie superalgebra, but an $F-$Lie
algebra. The $F-$Lie algebras were introduced in \cite{flie1, flie2}. They
admit a $\ZZ_F-$gradation, the zero-graded part
being a Lie algebra
and the non-zero graded parts being appropriate representations of
the zero-graded part. An $F-$fold symmetric product (playing the role of
the anticommutator in the case $F=2$)  expresses the zero graded  part
in terms of the non-zero graded part. In particular when the zero-graded
part coincides with the Poincar\'e algebra, one has additional generators,
denoted $V$, such that  $V^F \sim P$,  with $P$ the 
space-time translation generators.  Then, it has been realised that,
within the framework of $F-$Lie algebras, different extensions 
of the Poincar\'e algebra can be
constructed.\\ 
Firstly,  one considers  the new generators $V$ belonging to an
infinite dimensional representation of the Poincar\'e algebra (a
Verma module) \cite{flie1}.
Such extensions  in  $(1+2)-$dimensional space-time 
have been considered, with $V$ belonging to the 
spin $1/F-$representation of the  $(1+2)D$ Poincar\'e group.  
It has been shown that 
this symmetry acts on relativistic anyons \cite{3d} and  can be seen
as a direct generalisation of supersymmetry, named fractional supersymmetry
\cite{fsusy}.  \\
Secondly,  finite dimensional $F-$Lie algebras have been  obtained by an
inductive theorem, starting from any Lie (super)algebras \cite{flie2}.
In\"on\"u-Wigner contractions of certain of them allow
other types of extensions of the Poincar\'e algebra.

Following this second line,
in a previous paper \cite{cubic} a field theoretical 
realisation of the simplest
extension (with $F=3$) has been realised in
four space-time dimensions. 
This new extension of the Poincar\'e algebra
was named cubic supersymmetry or 3SUSY algebra. Since this extension
is based on some new generators in the vector representation of the
Poincar\'e algebra
(see bellow), its representations contain states of a definite
statistics, {\it i.e.} the multiplets 
are purely bosonic or purely fermionic. Thus, the
name cubic supersymmetry  may be misleading. In order to avoid
confusion, from now on, this symmetry will be called cubic symmetry.

In this paper, we  will extend to 
an arbitrary number of spacetime dimensions some of the results
obtained in \cite{cubic}. Studying the representation
theory of this algebra, one obtains, among other possibilities,
antisymmetric tensor multiplets  containing    $p-$forms.
We thus have a symmetry  that  acts naturally on these $p-$forms.

In the next section, we recall briefly the underlying algebra 
and its matrix representation in $D-$space-time dimensions.
Then, we explicitly consider the antisymmetric tensor multiplets
and their transformation laws in $D=1+(4n-1)$. 
These transformations acting on $p-$forms have an interesting 
geometrical interpretation in terms of the inner and exterior 
product with a one-form, the parameter of the transformation.
In section 3, invariant Lagrangians  are obtained. We observe that
 cubic symmetry invariance requires gauge fixing
terms adapted for $p-$forms  thus generalising the  usual 
gauge fixing term of electrodynamics. In the last section some conclusions
are given.

\section{Non-trivial extension of the Poincar\'e algebra}

In this section we will extend to an arbitrary number of spacetime dimensions
the cubic symmetry algebra and its matrix representation.
Antisymmetric tensor multiplets and their transformations laws
will then be obtained.

\subsection{The algebra and matrix representations}

The cubic symmetry algebra  is constructed from the Poincar\'e generators
($L_{MN}$ and $P_M$, $M,N=0,\ldots, D-1$) together 
with some  additional generators 
$V_M$ in the vector representation of the Poincar\'e algebra,

 \beqa
\label{algebra}
&&\left[L_{MN}, L_{PQ}\right]=
\eta_{NP} L_{PM}-\eta_{MP} L_{PN} + \eta_{NP}L_{MQ}-\eta_{MP} L_{NQ},
\nonumber \\
 &&\left[L_{MN}, P_P \right]= \eta_{NP} P_M -\eta_{MP} P_N, \nonumber \\
&&\left[L_{MN}, V_P \right]= \eta_{NP} V_M -\eta_{MP} V_N, \ \
\left[P_{M}, V_N \right]= 0, \\
&&\left\{V_M, V_N, V_R \right \}=
\eta_{MN} P_R +  \eta_{MR} P_N + \eta_{RN} P_M, \nonumber
\eeqa

 \noindent
where $\{V_M,V_N,V_P \}=
V_M V_N V_R + V_M V_R V_N + V_N V_M V_R + V_N V_R V_M + V_R V_M V_N +
V_R V_N V_M $  stands for the symmetric product of order $3$ and
$\eta_{MN} = \mathrm{diag}\left(1,-1,\ldots ,-1\right)$  is the $D-$dimensional
 Minkowski
metric.
A matrix representation of our algebra is given by

\begin{eqnarray}
\label{matred}
V_M=\begin{pmatrix} 0&\Lambda^{1/3} \Gamma_M& 0 \cr
                           0&0&\Lambda^{1/3}\Gamma_M \cr
                           \Lambda^{-2/3}P_M&0&0\end{pmatrix}
\end{eqnarray}

\noindent
with $\Gamma_M$ the $D-$dimensional $\Gamma-$matrices
($\left\{\Gamma_M,\Gamma_N\right\}=2 \eta_{MN}$),
$P_M=\partial_M$\footnote{
It should be noted that, in this paper, we take 
 the structure constants to be real --see Eq.(\ref{algebra})--
and consequently there is no $i$ factor in $P_M$.}
 and $\Lambda$ a parameter with mass dimension.
When $D$  is an even number ($D=2k$) the $\Gamma-$matrices can be written
as $\Gamma_M = \begin{pmatrix} 0& \Sigma_M \cr \tilde \Sigma_M &0 \end{pmatrix}$,
with $\Sigma_0=\tilde \Sigma_0=1, \tilde \Sigma_I= -\Sigma_I, I=1,\ldots,
D-1$ and $\Sigma_I$ the generators of the Clifford algebra $SO(2k-1),
\Sigma_I \Sigma_J+ \Sigma_J \Sigma_I = 2 \delta_{IJ}$.  
Thus, the representation (\ref{matred})
is reducible leading to two inequivalent representations:

\begin{eqnarray}
\label{matirred}
V_{+}{}_M=\begin{pmatrix} 0&\Lambda^{1/3} \Sigma_M& 0 \cr
                           0&0&\Lambda^{1/3}\tilde \Sigma_M \cr
                           \Lambda^{-2/3}P_M&0&0 \end{pmatrix}, \nonumber \\
\\
V_{-}{}_M=\begin{pmatrix} 0&\Lambda^{1/3} \tilde \Sigma_M& 0 \cr
                           0&0&\Lambda^{1/3}\Sigma_M \cr
                           \Lambda^{-2/3}P_M&0&0 \end{pmatrix}. \nonumber 
\end{eqnarray} 
For simplicity, we set 
from now on  $\Lambda=1$
(in appropriate units). It has been noticed before that particles in an
irreducible representation of (\ref{algebra}) are degenerate in mass 
\cite{cubic}, because $P^2$ is a Casimir operator.

For further use we introduce the antisymmetric set of $\Gamma$ matrices
\begin{eqnarray}
\label{gamma2}
\Gamma^{(\ell)}: \Gamma_{M_1 \cdots M_\ell} =
\frac{1}{\ell !} \sum \limits_{\sigma \in S_\ell} 
\Gamma_{M_{\sigma(1)} }\cdots  
\Gamma_{M_{\sigma(\ell )} }   
\end{eqnarray}

\noindent 
(with $S_\ell$ the group of permutations with $\ell$ elements)
which for even $D$ gives a further simplification

\begin{eqnarray}
\label{gamma}
\Gamma_{ M_1 \cdots M_{2 \ell}  }& =& \begin{pmatrix}
\Sigma_{M_1} \tilde \Sigma_{M_2} \cdots \Sigma_{M_{2 \ell-1}} 
\tilde \Sigma_{M_{2 \ell}} 
+\mathrm{~perm}&0 \cr
0&  \tilde  \Sigma_{M_1}  \Sigma_{M_2} \cdots \tilde 
\Sigma_{M_{2 \ell-1}} \Sigma_{M_{2 \ell}}
+\mathrm{~perm}
 \end{pmatrix}
\nonumber \\
&=&  \begin{pmatrix}  \Sigma_{{M_1} \cdots M_{2 \ell}}&0 \cr
0&\tilde  \Sigma_{{M_1} \cdots M_{2 \ell}} \end{pmatrix} 
\nonumber \\ 
\\ \nonumber \\
 \Gamma_{ M_1 \cdots M_{2 \ell + 1 }  }& =& \begin{pmatrix} 
0& \Sigma_{M_1} \tilde \Sigma_{M_2} \cdots  \tilde 
\Sigma_{M_{2 \ell  } }\Sigma_{M_{2 \ell + 1 }}
+\mathrm{~perm}\cr
  \tilde  \Sigma_{M_1}  \Sigma_{M_2} \cdots  \Sigma_{M_{2 \ell  }}
\tilde  \Sigma_{M_{2 \ell + 1}}
+\mathrm{~perm}&0
 \end{pmatrix} \nonumber \\ 
&=&\begin{pmatrix}  0 &  \Sigma_{{M_1} \cdots M_{2 \ell +1 }} \cr
\tilde  \Sigma_{{M_1} \cdots M_{2 \ell + 1}} &0 \end{pmatrix}.
\nonumber
\end{eqnarray} 

\noi 
where the definitions of the $\Sigma, \tilde \Sigma$ matrices can be deduced 
from the equalities above.
  For instance,
 $\Sigma_{M_1 \ldots M_{2\ell}} =
\sum \limits_{\sigma \in  S_\ell} 
\Big( \prod \limits_{i=1}^\ell \Sigma_{M_{\sigma(2i-1)}}  
\tilde \Sigma_{M_{\sigma(2 i)}} \Big)$
and similarly for the other matrices.

\subsection{Antisymmetric  tensor multiplets}
To build a representation of the algebra (\ref{algebra})
using the matrix representations (\ref{matred}) and (\ref{matirred})
 we have also
to specify the representation of the vacuum. If the vacuum is in the trivial 
representation
of the Lorentz algebra, the multiplet  consists of 
three spinors $\mathbf{\Psi}=
\begin{pmatrix} \Psi_{ 1 } \cr
 \Psi_2{} \cr \Psi_{3 }\end{pmatrix}$  transforming as 
 $\delta_\e {\mathbf{\Psi}} = 
\e^M V_M \mathbf{\Psi}$. When the space-time dimension is
even, this representation is reducible and consist of
three spinors 
of definite  chirality. We have two possibilities
corresponding to the two choices for the matrices  
$V_\pm$~(see (\ref{matirred}))
$\mathbf{\Psi_+}=
\begin{pmatrix} \Psi_{ 1}{}_+ \cr
\bar \Psi_2{}_{-} \cr \Psi_{3}{}_+ \end{pmatrix}$  or
$\mathbf{\Psi_-}=
\begin{pmatrix}  \bar \Psi_{ 1}{}_{-} \cr
 \Psi_2{}_{+} \cr \bar \Psi_{3}{}_{-} \end{pmatrix}$. They
transform as $\delta_\e {\mathbf{\Psi}}_\pm = 
\e^M V_\pm{}_M \mathbf{\Psi}_\pm$, with $\e$ a 
{\it commuting Lorentz vector}
that we take real
(we stress that $\e$ is not an anticommuting spinor unlike in supersymmetry),
 $\Psi_+$ a left-handed spinor and $\bar \Psi_{-}$ 
a right-handed spinor. 

>From now on we consider the case of $D=1+(4n-1)$ space-time dimensions,
the general case will be studied elsewhere.
We concentrate on  the case where the vacua 
are in the spinor representations of the Lorentz algebra.
We take two copies $\mathbf{\Psi}_\pm, \mathbf{\Lambda}_\pm$ transforming
with $V_\pm$, and two copies of the vacuum in the spinor  representation
$\Omega_\pm, \omega_\pm$. From the decomposition of the product
of spinors on the set of $p-$forms

\newpage
\beqa
\label{bispin}
{\cal S}_+ \otimes {\cal S}_+ &=&
 [0] \oplus [2] \oplus \cdots [2n]_+
\nonumber \\
{\cal S}_- \otimes {\cal S}_- &=&
 [0] \oplus [2] \oplus \cdots [2n]_- \\
{\cal S}_+ \otimes {\cal S}_- &=&  
 [1] \oplus [3] \oplus \cdots [2n-1] \nonumber 
\eeqa

\noindent
with ${\cal S}_\pm$ a left/right handed spinor and 
$[p]$ representing a $p-$form and $[2n]_\pm$ an (anti-)self-dual
$2n-$form,  one gets
the four multiplets

\beqa
\label{4-decomposition}
\Xi_{++}&=&\mathbf{\Psi_+} \otimes \Omega_+ =
\begin{pmatrix} \Xi_1{}_{++} \cr 
 \bar \Xi_2{}_{-+} \cr \Xi_3{}_{++} 
\end{pmatrix} = \begin{pmatrix}
A_{[0]}\oplus  A_{[2]}\oplus   \cdots\oplus A_{[2n]_+} \cr
\tA_{[1]}\oplus\tA_{[3]}\oplus\cdots\oplus\tA_{[2n-1]} \cr
\ttA_{[0]} \oplus\ttA_{[2]} \oplus\cdots \oplus\ttA_{[2n]_+}
\end{pmatrix}
\nonumber \\
\Xi_{--}&=&\mathbf{\Psi_-} \otimes \Omega_- =
\begin{pmatrix} \bar \Xi_1{}_{--} \cr 
 \Xi_2{}_{+-} \cr \bar 
\Xi_3{}_{--} 
\end{pmatrix}
=\begin{pmatrix}
A'_{[0]}\oplus A'_{[2]} \oplus\cdots \oplus A'_{[2n]_-} \cr
\tA'_{[1]}\oplus\tA'_{[3]}\oplus\cdots\oplus\tA'_{[2n-1]} \cr
\ttA'_{[0]} \oplus\ttA'_{[2]} \oplus\cdots \oplus\ttA'_{[2n]_-}
 \end{pmatrix}
\nonumber \\
\\
\Xi_{-+}&=&\mathbf{\Lambda_-} \otimes \omega_+ =
\begin{pmatrix} \xi_1{}_{-+} \cr 
 \bar \xi_2{}_{++} \cr \xi_3{}_{-+} 
 \end{pmatrix}=
\begin{pmatrix}
A_{[1]} \oplus A_{[3]} \oplus\cdots\oplus A_{[2n-1]} \cr
\tA_{[0]}\oplus\tA_{[2]}\oplus\cdots,\tA_{[2n]_+} \cr
\ttA_{[1]}\oplus \ttA_{[3]} \oplus\cdots \oplus\ttA_{[2n-1]}
  \end{pmatrix}
\nonumber \\
\Xi_{+-}&=&\mathbf{\Lambda_+} \otimes \omega_- =
\begin{pmatrix} \bar \xi_1{}_{+-} \cr 
 \xi_2{}_{--} \cr \bar 
\xi_3{}_{ +-} 
\end{pmatrix} =
  \begin{pmatrix}
A'_{[1]}\oplus A'_{[3]} \oplus\cdots\oplus A'_{[2n-1]} \cr
\tA'_{[0]}\oplus\tA'_{[2]}\oplus\cdots\oplus\tA'_{[2n]_-} \cr
\ttA'_{[1]}\oplus \ttA'_{[3]} \oplus\cdots \oplus\ttA'_{[2n-1]}
\end{pmatrix}.
\nonumber 
\eeqa

\noindent 
Thus in each $\Xi$ multiplets we have various set of $p-$forms, with
$0\le p\le 2n$.
Due  to the property of (anti-)self-duality  of  $2n$-forms in
 $1+(4n-1)-$dimensions, ${}^\star A_{[2n]_+}= i A_{[2n]_+},  
{}^\star A'_{[2n]_-}= -i A'_{[2n]_-}$, {\it etc} 
(with $ {}^\star A_{[2n]_+} $ the
Hodge dual of $ A_{[2n]_+}$), the $2n$-forms are 
complex representations of
$SO(1,D-1)$ and consequently also the other $p-$forms
(see Eq.[\ref{transfo-form}] below). 
The multiplets in (\ref{4-decomposition}) can be taken complex
conjugate of each other ($\Xi_{++}^\star= \Xi_{--}$ and 
$\Xi_{+-}^\star= \Xi_{-+}$), that is

\beqa
\label{cc}
\begin{array}{ll}
A_{[2p]}^\star= A^\prime_{[2p]},& A_{[2n]_+}^\star=A^\prime_{[2n]_-} 
\\
{\tilde {\tilde A}}_{[2p]}^\star= 
{\tilde {\tilde A}}^\prime_{[2p]},& 
{\tilde {\tilde A}}_{[2n]_+}^\star={\tilde {\tilde A}}^\prime_{[2n]_-} \\
\tilde A^\star_{[2p+1]}=A^\prime_{[2p+1]} &
\end{array}
\eeqa

\noindent
(the complex conjugate $A^\star$ of $A$ should not to be confused
with its dual ${}^\star A)$
and similarly for the fields coming from $\Xi_{-+}$ and $\Xi_{+-}$.

The underlying algebra (\ref{algebra}) and its representations
(\ref{matred}) and (\ref{matirred})   have a $\ZZ_3-$graded structure.
Hence, one can assume that the fields with no tilde symbol are in the
$(-1)-$graded sector, the fields with one tilde are in the  $0$-graded
sector and the fields with two tilde symbols are in the $1-$graded
sector. For example, for the multiplet $\Xi_{++}$, the fields
$A_{[0]}, \ldots, A_{[2n]_+}$ are in the $(-1)-$graded sector,
the fields $\tA_{[1]},\ldots,\tA_{[2n-1]}$ are in the $0-$graded
sector and the fields $\ttA_{[0]}, \ldots, \ttA_{[2n]_+}$ 
are in the $1-$graded sector.

Now, in order to obtain the transformation laws of the $p-$forms,
we give first the general relations which allow to project out  
any $p-$form $\omega_{[p]}$
from the product of two spinors, denoted 
$\Psi_\pm \otimes \Psi_\pm$. One has 

\beqa
\label{spin-p}
\Psi_{++}= \Psi_+ \otimes  \Psi^\prime{}^t_+{\cal C}_+&=&
 \sum \limits_{p=0}^{n-1} \frac{1}{(2p)!} \omega_{[2 p]} \Sigma^{(2p)}
+ \frac12\frac{1}{(2n)!} \omega_{[2n]_+} \Sigma^{(2n)} \nonumber \\
\Psi_{--}= \Psi_- \otimes  \Psi^\prime{}^t_-{\cal C}_-&=&
 \sum \limits_{p=0}^{n-1} \frac{1}{(2p)!} \omega^\prime{}_{[2 p]} 
\tilde \Sigma^{(2p)} 
+ \frac12\frac{1}{(2n )!} \omega^\prime{}_{[2n]_-} \tilde \Sigma^{(2n)} 
\nonumber \\ 
\\
\Psi_{+-}=\Psi_+ \otimes  \Psi^\prime{}^t_-{\cal C}_-&=&
 \sum \limits_{p=0}^{n-1} \frac{1}{(2p+1)!} \omega_{[2 p+1]} \Sigma^{(2p+1)} 
\nonumber \\
\Psi_{-+}=\Psi_- \otimes \Psi^\prime{}^t_+ {\cal C}_+&=&
 \sum \limits_{p=0}^{n-1} \frac{1}{(2p+1)!} \omega^\prime{}_{[2 p+1]} \tilde 
\Sigma^{(2p+1)}.
\nonumber 
\eeqa

\noindent
Since the $\Sigma$ matrices (\ref{gamma}) act on spinors, they have the first
spinor index up and the second  spinor index down. This means that  
in the   equations above the index of the second spinor has to be raised
by means of the charge conjugation   matrix 
${\cal C}=\begin{pmatrix} {\cal C}_+&0 \cr 0 & {\cal C}_- \end{pmatrix}$. 
In the correspondence (\ref{spin-p}), we 
use  the symbolic notations  $\omega_{[2 p]} \Sigma^{(2p)} =
\omega_{[2p]}{}_{M_1 \cdots M_{2p}} \Sigma^{M_{2p} \cdots M_1}$.
These relations which originate from  the
properties of the Dirac $\Gamma-$matrices, translate in fact (\ref{bispin})
with explicit normalisations.
Conversely, using the trace properties of the $\Gamma$ matrices one gets

\beqa
\label{p-spin2}
\omega_{[2p]}&=&  \frac{1}{2^{2n}} 
{\mathrm{Tr}}\left(\Sigma^{(2p)} \Psi_{++}\right),
\ \ 
\omega_{[2n]_+}{} = 
\frac{1}{2^{2n}} {\mathrm{Tr}}\left(\Sigma^{(2n)} \Psi_{++}
\right),
\nonumber \\
\omega^\prime{}_{[2p]}&=&  \frac{1}{2^{2n}} 
{\mathrm{Tr}}\left(\tilde \Sigma^{{(2p)}}  
\Psi_{--}\right),\ \ 
\omega{}^\prime_{[2n]_-} = \frac{1}{2^{2n}} 
{\mathrm{Tr}}\left(\tilde \Sigma^{(2n)} 
\Psi_{--}\right), \nonumber \\
\omega_{[2p+1]}&=&  \frac{1}{2^{2n}} \mathrm{Tr}\left(\tilde \Sigma^{{(2p+1)}} 
\Psi_{+-}\right),
\\
\omega^\prime{}_{[2p+1]}&=&  \frac{1}{2^{2n}} 
\mathrm{Tr}\left(\Sigma^{{(2p+1)}} 
\Psi_{-+}\right).
\nonumber 
\eeqa

\noindent
in components this gives for instance 
$\omega_{[2p]}{}^{M_1 \ldots M_{2p}}=  \frac{1}{2^{2n}} 
\mathrm{Tr}\left(\Sigma^{M_1 \ldots M_{2p}} \Psi_{++}\right)$.
Using the relations (\ref{spin-p}) and (\ref{p-spin2}) can 
associate the  $p-$forms  listed in 
(\ref{4-decomposition}) to the multiplets
$\Xi_{\pm \pm}$.
We now  calculate the transformation laws of these various $p-$forms.
For example, for  $\Xi_{++}$, from the transformation 

\beqa
\label{transfo++}
\delta_\e \Xi_{++}= \left(\e^M V_M {\mathbf \Psi}_+\right) \otimes 
{\mathbf \Omega}_+
\eeqa
\noindent
we obtain

\beqa
\label{transfo-spin}
\delta_\e \Xi_1{}_{++} &=& \e^M \Sigma_M \bar \Xi_2{}_{-+}, \nonumber \\
\delta_\e \bar \Xi_2{}_{-+} &=& \e^M \tilde \Sigma_M \Xi_3{}_{++}, \\
\delta_\e \Xi_3{}_{++} &=& \e^M \partial_M \Xi_1{}_{++}. \nonumber 
\eeqa

\noindent
To calculate from (\ref{transfo-spin}) the transformation laws
of the  $p-$forms, we use the identities 

\beqa
\label{identities}
\Sigma_{M_1 \ldots M_{2k}} \tilde \Sigma_{M_{2k+1}}=
\Sigma_{M_1 \ldots M_{2k} M_{2k+1}} +
 \eta_{M_{2k} M_{2k+1}}\Sigma_{M_1 \ldots M_{2k-1}} + {\mathrm perm.}
\nonumber \\
\Sigma_{M_1 \ldots M_{2k+1}}  \Sigma_{M_{2k+2}}=
\Sigma_{M_1 \ldots M_{2k+1} M_{2k+2}} +
 \eta_{M_{2k+1} M_{2k+2}}\Sigma_{M_1 \ldots M_{2k}} + {\mathrm perm.}
\eeqa

\noindent 
(where perm means sum on all permutations with the sign corresponding 
to its signature)
for $k <n$ (when $ k\ge  n$ the situation is more involved because
the $\Sigma^{(2 k+2)}, \tilde \Sigma^{(2k+1)}$ matrices are related to 
the $\Sigma^{(2k+2-2n)}, \tilde \Sigma^{(2k+1 - 2n)}$ matrices).
Similar relations hold for the $\tilde \Sigma^{(k)}$ matrices.

To obtain from (\ref{transfo-spin}) the transformation properties
of the $p-$forms, we proceed differently for $\Xi_1{}_{++}$ and
$\bar \Xi_2{}_{-+}$, in order to avoid the presence of $\Sigma^{(\ell)}$
or $\tilde \Sigma^{(\ell)}$ matrices with $\ell >2n$.

Starting from  $\delta_\e \bar \Xi_2{}_{-+} = \e^M \tilde \Sigma_M 
\Xi_3{}_{++}$, and 
using (\ref{p-spin2}) we first have 
$\delta_\e \tA_{[2p+1]}=  \e^M \frac{1}{2^{2n}} \mathrm{Tr} 
\left( \Sigma^{(2p+1)} \tilde \Sigma_M \Xi_3{}_{++}\right)$. 
Using (\ref{identities}), 
 we calculate $ \Sigma^{(2p+1)} \tilde \Sigma_M$. Then, from the trace 
identities for the $\Sigma$ matrices, one
obtains the transformations laws for $\tA_{[2p+1]}$ (see (\ref{transfo-form})).

In order to calculate  $\delta_\e \Xi_1{}_{++} = 
\e^M  \Sigma_M \bar \Xi_2{}_{-+}$,
we proceed in the reverse order. Firstly, using (\ref{spin-p}) and
the identity (\ref{identities}), we compute 
the product $\Sigma_M \bar \Xi_2{}_{-+}$.
 Then,   using the trace formulae, we  get the transformation laws of
$A_{[2p]}$. 

The last case $\delta_\e \Xi_3{}_{++}$ is not
difficult to handle. Finally we get

\beqa
\label{transfo-form}
&&\begin{array}{llllll}
\delta_\e A_{[0]}&=& i_\e {\tilde A}_{[1]}&
\delta_\e {\tilde A}_{[1]}&=& i_\e {\tilde {\tilde A}}_{[2]}
+ \ttA_{[0]} \wedge \e \cr
&\vdots && &\vdots \cr
\delta_\e A_{[2p]}&=& i_\e {\tilde A}_{[2p+1]} + {\tilde A}_{[2p-1]} 
\wedge \e &
\delta_\e {\tilde A}_{[2p+1]}&=& i_\e {\tilde {\tilde A}}_{[2p+2]} + {\tilde 
{\tilde A}}_{[2p]} \wedge \e  \cr
&   \vdots && &  \vdots \cr
\delta_\e A_{[2n]_+}&=&  {\tilde A}_{[2n-1]} \wedge \e
-i
 \ {}^\star \hskip -.1truecm
\left
(\tilde A_{[2n-1]} \wedge \e\right)
 &
\delta_\e {\tilde A}_{[2n-1]}&=& i_\e {\tilde {\tilde A}}_{[2n]_+}+
{\tilde {\tilde A}}_{[2n-2]} \wedge \e 
 \cr
\end{array} \nonumber \\
\\
&&\ \ \delta_\e {\tilde {\tilde A}}_{[0]}= \e A_{[0]}, \ \ \ \ \ \ldots
 \ \ \ \ \ 
\delta_\e {\tilde {\tilde A}}_{[2n]_+}= \e A_{[2n]_+}. \nonumber 
\eeqa

\noindent
In these relations $i_\e A$ represents the inner product of $\e$ and
$A$\footnote{In this paper, we take
$(i_\varepsilon A_{[p]})_{M_1 \cdots M_{p-1}}= A_{[p]}{}_{M_1 \cdots M_p}
\varepsilon^{M_p}$.},   $A \wedge \e$ represents 
the exterior product of $A$ and $\e$, 
 $\e A$ represents the action of the vector field $\e = \e^M \partial_M$
on $A$ and  ${}^\star A$ represents the dual of 
$A$\footnote{We remark that the  transformation laws 
(\ref{transfo-form}) have a geometrical interpretation in
terms of inner and exterior product.}. 
The terms $-i
 \ {}^\star \hskip -.1truecm
\left
(\tilde A_{[2n-1]} \wedge \e\right)$ in $\delta_\e A_{[2n]_+}$ preserve
the  self-dual character of $A_{[2n]_+}$.  Similar transformation laws
can be obtained for the other multiplets. 
In the case of multiplets
involving anti-self-dual $2n-$forms, the $-i$ becomes a $+i$, 
in perfect agreement with the complex conjugation (\ref{cc}).

\section{Invariant Lagrangians}
The transformations (\ref{transfo-form}) suggest that an invariant
Lagrangian should contain only zero-graded terms.
 Thus, one is forced
to couple the fields in the $(-1)-$graded sector to the fields in
the $1-$graded sector and the fields in the $0-$graded sector to themselves.
Furthermore, if we consider for example the $\Xi_{++}$ multiplet,
in order to have a real Lagrangian,  one
has  also to take into consideration
the conjugate multiplet $\Xi_{--}$ (see (\ref{cc})).
For the $\Xi_{++}$ and $\Xi_{--}$ multiplets, the Lagrangian writes

\beqa
\label{lag}
{\cal L}&=& {\cal L}(\Xi_{++}) + {\cal L}(\Xi_{--}) =
{\cal L}_{[0]} + \ldots + {\cal L}_{[2n]} 
 + {\cal L}'_{[0]} \ \ + \ldots + {\cal L}'_{[2n]} 
 \nonumber \\
&=&  
 d A_{[0]} d {\tilde {\tilde A}}_{[0]} 
+ \ldots + \nonumber \\
&-& 
\frac12  \frac{1}{(2p+2)!} d \tilde A_{[2p+1]} d \tilde A_{[2p+1]} 
-\frac12  \frac{1}{(2p)!} d^\dag \tilde A_{[2p+1]} d^\dag \tilde A_{[2p+1]} 
 \nonumber \\ 
&+& \frac{1}{(2p+3)!} d A_{[2p+2]} 
d {\tilde {\tilde A}}_{[2p + 3]} 
+ \frac{1}{(2p+1)!} d^\dag A_{[2p+2]} d^\dag {\tilde { \tilde A}}_{[2p +2]} 
 \nonumber  \\
&+& \ldots  + \nonumber \\
&+&\frac12 \frac{1}{(2n+1)!} d A_{[2n]_+} 
d {\tilde {\tilde A}}_{[2n]_+} 
+\frac12
 \frac{1}{(2n-1)!} d^\dag A_{[2n]_+} d^\dag {\tilde { \tilde A}}_{[2n]_+} 
 \nonumber \\
&+&  d A'_{[0]} d {\tilde {\tilde A}}'_{[0]} 
+ \ldots +  \\
&-& 
\frac12  \frac{1}{(2p+2)!} d \tilde A'_{[2p+1]} d \tilde A'_{[2p+1]} 
- \frac12  \frac{1}{(2p)!} d^\dag \tilde A'_{[2p+1]} d^\dag \tilde A'_{[2p+1]} 
 \nonumber \\
 &+& \frac{1}{(2p+3)!} d A'_{[2p+2]} 
d {\tilde {\tilde A}}'_{[2p + 3]} 
+ \frac{1}{(2p+1)!} d^\dag A'_{[2p+2]} d^\dag {\tilde { \tilde A}}'_{[2p +2]} 
 \nonumber  \\ 
&+& \ldots  + \nonumber \\
&+&\frac12 \frac{1}{(2n+1)!} d A'_{[2n]_-} 
d {\tilde {\tilde A}}'_{[2n]_-} 
+ \frac12 \frac{1}{(2n-1)!} d^\dag A'_{[2n]_-} d^\dag {\tilde 
{ \tilde A}}'_{[2n]_-}.
 \nonumber
\eeqa

\noindent
Here $\omega_{[p]} \omega'_{[p]}$ stands for
$\omega_{[p]}{}_{M_1\ldots M_p} 
$\ $\omega'_{[p]}{}^{M_1 \ldots M_P}$
, where $\omega_{[p]}$
and $\omega'_{[p]}$ are two $p-$forms~\footnote{In $p-$form
notation  we can rewrite $\int d^Dx~ \omega_{[p]} \omega'_{[p]}$
as $\int \omega_{[p]} \wedge {}^\star \omega'_{[p]}$.}.
In the Lagrangian (\ref{lag}), $d A_{[p]}$ is the exterior derivative of
$A_{[p]}$ and  $d^\dag A_{[p]}$ its adjoint
$d^\dag A_{[p]}= (-1)^{pD+D} {}^\star d {}^\star A_{[p]}
={}^\star d {}^\star A_{[p]}$   for even $D$.

To prove that (\ref{lag}) is invariant under (\ref{transfo-form}), we
firstly note that $\delta_\e {\cal L}(\Xi_{++})$ and 
$\delta_\e{\cal L}(\Xi_{--})$ 
do not  mix. It is thus  sufficient to check separately their  invariance,
which we do here only for ${\cal L}(\Xi_{++})$ as an illustration.
Starting from a specific normalisation for ${\cal L}_{[0]}$,
its variation fixes the normalisation for ${\cal L}_{[1]}$.
By a step-by-step process, the normalisations for 
${\cal L}_{[p]}, 0\le p \le 2n$
 are
also fixed. At the very end, all the terms 
of $\delta_\e {\cal L}$
compensate each others, up to a
total derivative. Thus, the Lagrangian (\ref{lag}) is invariant. \\

If one considers the terms involving the 
(anti-)self-dual $2n-$form one can have   further simplifications.
Indeed, for the self-dual $2n-$form we have
$ {\cal L}_{[2n]} =\frac12 \frac{1}{(2n+1)!} d A_{[2n]_+} 
 d {\tilde {\tilde A}}_{[2n]_+} + \frac12
 \frac{1}{(2n-1)!} d^\dag A_{[2n]_+} d^\dag {\tilde {\tilde A}}_{[2n]_+}=
\frac{1}{(2n+1)!} d A_{[2n]_+}   
d {\tilde {\tilde A}}_{[2n]_+}$
because of the self-duality condition 
${}^\star A_{[2n]_+}=  i A_{[2n]_+}$.
A more interesting way of regrouping the  terms involving the
self-dual and the anti-self-dual $2n-$forms  is 

\beqa
\label{2n-form}
{\cal L}_{[2n]} + {\cal L}'_{[2n]}& =& 
\frac12 \frac{1}{(2n+1)!} d A_{[2n]_+} 
d {\tilde {\tilde A}}_{[2n]_+} 
+\frac12
 \frac{1}{(2n-1)!} d^\dag A_{[2n]_+} d^\dag {\tilde { \tilde A}}_{[2n]_+} 
 \nonumber \\
&+&\frac12 \frac{1}{(2n+1)!} d A'_{[2n]_-} 
d {\tilde {\tilde A}}'_{[2n]_-} 
+ \frac12 \frac{1}{(2n-1)!} d^\dag A'_{[2n]_-} d^\dag {\tilde 
{ \tilde A}}'_{[2n]_-} \nonumber \\
&=&\frac12 \frac{1}{(2n+1)!} d \big(A_{[2n]_+}+ A'_{[2n]_-} \big)
d \big(\ttA_{[2n]_+}+ \ttA'_{[2n]_-} \big)  \\
&+ &\frac12 \frac{1}{(2n-1)!} d^\dag \big(A_{[2n]_+}+A'_{[2n]_-} \big)
d^\dag \big(\ttA_{[2n]_+}+\ttA'_{[2n]_-} \big),
 \nonumber
\eeqa

\noindent
since $A_{[2n]_+} A'_{[2n]_-} =0$ when $D=1+(4n-1)$, for a self-dual and
an anti-self-dual $2n-$forms. The final real $2n-$forms 

\beqa
\label{2n-real}
A_1{}_{[2n]}= \frac{1}{\sqrt{2}}\left(A_{[2n]_+} +A'_{[2n]_-}\right),
\ttA_1{}_{[2n]}= \frac{1}{\sqrt{2}}\left(\ttA_{[2n]_+} +\ttA'_{[2n]_-}\right),
\eeqa

\noi
are neither self-dual nor anti-self-dual, which is in 
agreement with the  representation  theory of the Poincar\'e algebra.

Considering the gauge invariance
\beqa
\label{gauge}
A_{[p]}\rightarrow A_{[p]}+d\chi_{[p-1]}, \mbox{ with } p\ge 1,
\eeqa
(where $\chi_{[p-1]}$ is a $(p-1)-$form), the terms involving $d^\dag$
in ${\cal L}$ (\ref{lag}) fix partially the gauge
\beqa
\label{gf}
d^\dag d \chi_{[p-1]} = 0.
\eeqa
Hence, these terms involving the $d^\dag$ operators can be seen as 
gauge fixing terms (Feynman gauge adapted for $p-$forms).
Another way of seeing this is to rewrite the Lagrangian using Fermi-like
terms. For instance,  for the ${\cal L}_{[2p+1]}$ part in (\ref{lag}) we have 

\beqa
\label{fermi}
{\cal L}_{[2p+1]}&=& -\frac12 \left(\frac{1}{ (2p+2)!} d \tilde A_{[2p+1]}  
d \tilde A_{[2p +1 ]}  +  \frac{1}{ (2p)!} d^\dag \tilde A_{[2p+1]}  
d^\dag \tilde A_{[2p+1]}  \right)
\nonumber \\
&=& -\frac12 \frac{1}{ (2p+1) !} \partial_{M_1} \tilde A_{[2p+1]}{}_{M_2 
\cdots M_{2p+2}} \partial^{M_1} \tilde A_{[2p+1]}{}^{M_2  \cdots M_{2p+2}}
\nonumber \\
&=&-\frac12 \frac{1}{ (2p+1) !} \partial \tA_{[2p+1]} \partial \tA_{[2p+1]}
 \nonumber 
\eeqa

\bigskip

We now look at the physical degrees of freedom
in (\ref{lag}). By construction, the various $p-$forms are complex;
thus we introduce  real $p-$forms  through

\beqa
\label{real}
\begin{array}{ll}
A_1{}_{[2p]}=\frac{1}{\sqrt{2}}\left(A_{[2p]} + A'_{[2p]}\right),&
A_2{}_{[2p]}= \frac{i}{\sqrt{2}}\left(A_{[2p]} - A'_{[2p]}\right),
\cr
\tA_1{}_{[2p+1]}=\frac{1}{\sqrt{2}}
\left(\tA_{[2p+1]} + \tA'_{[2p+1]}\right),&
\tA_2{}_{[2p+1]}=\frac{i}
{\sqrt{2}}\left(\tA_{[2p+1]} - \tA'_{[2p+1]}\right),
\cr
\ttA_1{}_{[2p]}=\frac{1}{\sqrt{2}}\left(\ttA_{[2p]} + \ttA'_{[2p]}\right),&
\ttA_2{}_{[2p]}= \frac{i}{\sqrt{2}}
\left(\ttA_{[2p]} - \ttA'_{[2p]}\right).
\end{array}
\eeqa

\noindent
with $p=0,\ldots, n-1$.
Observe that in the substitution (\ref{2n-real})
we have two $2n-$forms, although for the other  substitution (\ref{real})
we have four $p-$forms. 
After these redefinitions, we have a Lagrangian expressed
only with real fields. However, some of the terms in ${\cal L}$ 
are not diagonal, namely the ones involving the fields 
of the $(-1)-$ and $1-$graded sectors ({\it e.g} 
$\frac{1}{(2p+1)!} d A_1{}_{[2p]} d \ttA_1{}_{[2p]}
+\frac{1}{(2p-1)!} d^\dag A_1{}_{[2p]} d^\dag \ttA_1{}_{[2p]}$).
Therefore, in order to diagonalise ${\cal L}$ we introduce

\beqa
\label{field-diag}
&&\hA_1{}_{[2p]} =\frac{1}{\sqrt{2}}\left(A_1{}_{[2p]} + 
\ttA_1{}_{[2p]}\right),
\hhA_1{}_{[2p]}= \frac{1}{\sqrt{2}}\left(A_1{}_{[2p]} - \ttA_1{}_{[2p]}\right),
p=0,\ldots, n
 \\
&&\hA_2{}_{[2p]}= \frac{1}{\sqrt{2}}\left(A_2{}_{[2p]} + \ttA_2{}_{[2p]}\right),
\hhA_2{}_{[2p]}= \frac{1}{\sqrt{2}}\left(A_2{}_{[2p]} - \ttA_2{}_{[2p]}\right),
p=0,\ldots, n-1, \nonumber 
\eeqa

\noindent
which are mixtures of fields of gradation $(-1)$ and $1$.\\

Finally, after the field redefinitions (\ref{real}) and
(\ref{field-diag}),
the new fields are $\hA_1{}_{[2p]}, \hA_2{}_{[2p]},$  
$\hhA_1{}_{[2p]}, \hhA_2{}_{[2p]}, \tA_1{}_{[2p+1]},\tA_2{}_{[2p+1]}, 
p=0,\ldots, n-1$ and $ \hA_1{}_{[2n]}, \hhA_1{}_{[2n]}$,
Usually, the kinetic term for a $p-$form  $\omega_{[p]}$
writes, with our convention for the metric, 
$(-1)^p \frac{1}{2(p+1)!} d \omega_{[p]} d \omega_{[p]}$.
Expressed with these fields, the Lagrangian has these
conventional kinetic  terms except that (i) we have also gauge fixing terms
in addition to the kinetic terms and (ii) some of the fields
($\hhA_1{}_{[2p]}, \hA_2{}_{[2p]},$ $ \tA_2{}_{[2p+1]}, p=0,\ldots, n-1,
\hhA_1{}_{[2n]}$), have the opposite sign for the kinetic and 
the gauge fixing terms. This implies that these fields have an
energy density not bounded from bellow. 

One possible way to avoid this problem is based on Hodge duality.
For a given $p-$form $\omega_{[p]}$ we have 

\beqa
\label{dual}
&(-1)^p&\left(\frac{1}{(p+1)!} d \omega_{[p]} d \omega_{[p]} +
\frac{1}{(p-1)!} d^\dag \omega_{[p]} d^\dag \omega_{[p]}\right) = \\
-&(-1)^{p}&
\left(\frac{1}{(D-p-1)!} d^\dag \rho_{[D-p]} d^\dag \rho_{[D-p]} 
+ \frac{1}{(D-p+1)!} d \rho_{[D-p]} d \rho_{[D-p]} \right) \nonumber 
\eeqa

\noindent
with $\rho = {}^\star \omega$ the Hodge dual of $\omega$.
This substitution is due to the special form of our Lagrangian:
through the substitution $\omega \to \rho$, the kinetic
term of $\omega$ becomes the gauge fixing term of $\rho$ and
{\it vice versa}. Thus the following substitutions  are made

\beqa
\label{dual-field}
\hhA_1{}_{[2p]} &\to& \hhB_1{}_{[D-2p]} =   {}^\star \hhA_1{}_{[2p]},
\nonumber \\
\hA_2{}_{[2p]} &\to& \hB_2{}_{[D-2p]} =  {}^\star \hA_2{}_{[2p]}, \nonumber \\
\tA_2{}_{[2p+1]} &\to& \tB_2{}_{[D-2p-1]} = {}^\star \tA_2{}_{[2p+1]} 
 \\
\hhA_1{}_{[2n]} &\to& \hhB_1{}_{[D-2n]}= {}^\star \hhA_1{}_{[2n]} 
\nonumber 
\eeqa

\noindent
with $p=0,\ldots, n-1$. 
Let us emphasize that  the substitutions (\ref{dual-field})
are done with respect to the gauge fields. This is quite
different from the usual duality transformations (generalising
the electric-magnetic duality) where the duality transformations
are done with respect to the field strengths. With the
``duality'' transformations (\ref{dual-field}) 
the number of degree of freedom
is not the same for $\omega_{[p]}$ and $\rho_{[D-p]} = {}^\star \omega_{[p]}$,
which is not the case for the usual duality transformations. 
Hence, the  two Lagrangians  (\ref{lag}) and (\ref{lag-end}) (see
below) describe inequivalent  theories.

Thus, starting from the Lagrangian (\ref{lag}) and performing
the field redefinitions (\ref{real}), (\ref{field-diag}) and 
(\ref{dual-field})
we get

\beqa
\label{lag-end}
&{\cal L}& = \frac12 d \hA_1{}_{[0]} d \hA_1{}_{[0]} +
\frac12 d \hhA_2{}_{[0]} d \hhA_2{}_{[0]}
\nonumber \\
 &+& \ldots +  \nonumber \\
&-&\frac12  \frac{1}{(2p+2)!} d \tilde A_1{}_{[2p+1]} d \tilde A_1{}_{[2p+1]}
-\frac12  \frac{1}{(2p)!} d^\dag \tilde A_1{}_{[2p+1]} 
d^\dag \tilde A_1{}_{[2p+1]} \nonumber \\
&+& \frac12 \frac{1}{(2p+3)!} d \hA_1{}_{[2p+2]} d \hA_1{}_{[2p+2]}
+ \frac12 \frac{1}{(2p+1)!} 
d^\dag\hA_1{}_{[2p+2]} d^\dag\hA_1{}_{[2p+2]} 
 \nonumber  \\
&+& \frac12 \frac{1}{(2p+3)!} d \hhA_2{}_{[2p+2]} d \hhA_2{}_{[2p+2]}
+ \frac12 \frac{1}{(2p+1)!} 
d^\dag\hhA_2{}_{[2p+2]} d^\dag\hhA_2{}_{[2p+2]} 
 \nonumber  \\
&+& \ldots +\nonumber\\
&+& \frac12 \frac{1}{(2n+1)!} d \hA_1{}_{[2n]} d \hA_1{}_{[2n]}
+ \frac12 \frac{1}{(2n-1)!} 
d^\dag\hA_1{}_{[2n-1]} d^\dag\hA_1{}_{[2n]} 
 \nonumber  \\
&+& \frac12 \frac{1}{(2n+1)!} d \hhB_1{}_{[2n]} d \hhB_1{}_{[2n]}
+ \frac12 \frac{1}{(2n-1)!} 
d^\dag\hhB_1{}_{[2n-1]} d^\dag\hhB_1{}_{[2n]} 
   \\
&+& \ldots +\nonumber\\
&+& \frac12 \frac{1}{(D-2p-1)!} d \hhB_1{}_{[D-2p-2]} d \hhB_1{}_{[D-2p-2]}
+ \frac12 \frac{1}{(D-2p-3)!} 
d^\dag\hhB_1{}_{[D-2p-2]} d^\dag\hhB_1{}_{[D-2p-2]} 
 \nonumber  \\
&+& \frac12 \frac{1}{(D-2p-1)!} d \hB_2{}_{[D-2p-2]} d \hB_2{}_{[D-2p-2]}
+ \frac12 \frac{1}{(D-2p-3)!} 
d^\dag\hB_2{}_{[D-2p-2]} d^\dag\hB_2{}_{[D-2p-2]} 
 \nonumber  \\
&-&\frac12  \frac{1}{(D-2p)!} d \tilde B_2{}_{[D-2p-1]} 
d \tilde B_2{}_{[D-2p-1]}
-\frac12  \frac{1}{(D-2p-2)!} d^\dag \tilde B_2{}_{[D-2p-1]} 
d^\dag \tilde B_2{}_{[D-2p-1]} \nonumber \\
&+& \ldots +\nonumber\\
&+&\frac12 \frac{1}{(D-1)!} d^\dag \hhB_1{}_{[D]} d^\dag \hhB_1{}_{[D]} 
+ \frac12 \frac{1}{(D-1)!} d^\dag \hB_2{}_{[D]} d^\dag \hB_2{}_{[D]}. 
\nonumber 
\eeqa

\noi
At the very end, we have one one-form, one  three-form, 
$\ldots$, one $(D-1)-$form in the zero-graded sector and 
two zero-forms, two two-forms, $\ldots$ and two $D-$form in the
mixture of the sectors of gradation $(-1)$ and $1$.
In the Lagrangian (\ref{lag-end}), all the $p-$forms have 
a kinetic term and a gauge fixing term; the only exceptions are 
the zero-forms,
which  have only  kinetic terms, and the $D-$forms, which
have only  gauge fixing terms.

The gauge invariance (\ref{gauge}) and the field equations imply
(for a $p-$form $A_{[p]}$, with $p \le D-2$) $P^M A_{[p]}{}_{MM_2\cdots M_p}=0$
and $P^2=0$ (with $P^M$ the momentum), thus $A_{[p]}$ gives rise
to a massless state in the  $p-$order  antisymmetric representation of
the little group $SO(D-2)$.  But, in our decomposition, there are  also
appearing $p-$forms with $p = D-1, D$. Of course these $p-$forms do not
propagate. It is interesting to note that such $p-$forms also appear
in $D=10$ in the context of type IIA, IIB string theory
\cite{pol}. 

As we have seen, cubic symmetry is 
compatible with  gauge invariance   if the gauge is partially fixed.
Thus, in order that the field equations  obtained from (\ref{lag-end})
reproduce the usual field equations for a free $p-$form, one has to impose
the analogous of the Lorentz condition $d^\dag A_{[p]}=0$
(to eliminate the unphysical components).
All this has to be implemented at the quantum level,
for instance using an appropriate extension of the Gupta-Bleuler
quantization for $p-$forms. Of course one has to check if this procedure is
compatible with cubic symmetry. 

Finally, a  massless $p-$form ($p < D-1$) has  $\begin{pmatrix} p \cr D-1 
\end{pmatrix}$ 
degrees of freedom off-shell and  $\begin{pmatrix} p \cr D-2 \end{pmatrix}$   on-shell,
a $(D-1)-$form has $1$ degree of freedom off-shell and no degree
of freedom on-shell and a $D-$form has no degree of freedom off- and on-shell.
Therefore, one can check that we have twice as many degrees of freedom in the
$(-1)-$ and $1-$graded sector than in the $0-$graded sector.

To the Lagrangian (\ref{lag}) one can add the following invariant mass terms

\beqa
\label{mass}
{\cal L}_m=m^2\Big(&& \hskip -.7truecm
A_{[0]} \ttA_{[0]} + \ldots -\frac12 \frac{1}{(2p+1)!}
\tA_{[2p+1]} \tA_{[2p+1]} + \frac{1}{(2p+2)!}A_{[2p+2]} \ttA_{[2p+2]} 
\nonumber \\
&&\hskip -.7truecm
+\ldots  +\frac{1}{(2n)!}A_{[2n]} \ttA_{[2n]}+
A'_{[0]} \ttA'_{[0]} + \ldots -\frac12 \frac{1}{(2p+1)!}
\tA'_{[2p+1]} \tA'_{[2p+1}  \nonumber  \\
&& \hskip -.7truecm +
\frac{1}{(2p+2)!}A'_{[2p+2]} \ttA'_{[2p+2]} +
\ldots  +\frac{1}{(2n)!}A'_{[2n]} \ttA'_{[2n]}\Big)
\eeqa

\noi
This means that with the same multiplets (\ref{4-decomposition})
one can have massless or massive invariant terms. Thus, the
matrix representations (\ref{matred}) and 
(\ref{matirred}) do not depend on whether or not we are considering
massless or massive particles.
Of course one can also proceed along the same lines as before 
(Eq.[\ref{real}]-[\ref{field-diag}]-[\ref{dual-field}]),
in order to express ${\cal L}_m$ in terms of the physical
degrees of freedom.

\section{Conclusion}
In this paper, we have explicitely built a $D-$dimensional field 
theory implementation of the cubic symmetry algebra. This symmetry acts 
naturally on $p-$forms and has a geometrical interpretation in terms of inner
and exterior products with a one-form, the parameter of the transformation.
Due to the interest for $p-$forms, mostly within supergravity and
superstring theories, it is appealing to consider new possible
symmetries on these antisymmetric tensor fields.
As we have seen, the cubic symmetry invariance requires the presence of
gauge fixing terms, thus placing the theory in a Feynman gauge adapted for 
$p-$forms. One of the possible invariant Lagrangians
contains   $p-$forms, with $0\le p \le D$. Amongst them, the $(D-1)-$ and 
$D-$forms are non-propagating, which is also the case for types IIA
and IIB string theories \cite{pol}. 
 Finally, when one considers the way the $p-$forms $A_{[2p]}$
 $\tA_{[2p+1]}$ transform under cubic symmetry (\ref{transfo-form}), 
one sees  an analogy with the $T-$duality transformations relating
the $p-$forms of types IIA and type IIB strings.  Indeed, 
looking at (\ref{transfo-form}) and taking 
$\e$ along a given direction, the transformation laws 
 correspond  to the $T-$duality formulae 
(C.1) and (C.2) of Ref.\cite{A-B} (when gravity is not present).

\bigskip
\noindent
{\bf Ackowledgement:} We would like to acknowledge J. Lukierski 
for useful discussions and
remarks.
\bigskip


\end{document}